\documentclass[11pt]{article}
\pdfoutput=1
\usepackage[utf8]{inputenc}

\usepackage[titletoc, toc, page]{appendix}
\usepackage[parfill]{parskip}
\usepackage{caption}
\usepackage{tabularx}
\usepackage{latexsym}
\usepackage{amstext,amsfonts,amssymb,amsmath}
\usepackage{dsfont}
\usepackage{subfig}
\usepackage{todonotes}
\usepackage{authblk}

\usepackage{multicol}
\usepackage{nonfloat}
\usepackage{color}
\usepackage{bm}
\usepackage[margin=1in]{geometry}
\usepackage{graphicx}
\usepackage{wrapfig}
\usepackage{enumitem}
\usepackage[font={small}]{caption}
\usepackage[citestyle=numeric,bibstyle=numeric,backend=biber,eprint=false,doi=false,isbn=false,url=false,sorting=none, maxbibnames=10,giveninits=true]{biblatex}
\renewbibmacro{in:}{}
\makeatletter
\def\blfootnote{\xdef\@thefnmark{}\@footnotetext}
\makeatother

\title{\huge Using machine learning to construct velocity fields from OH-PLIF images}
\date{}
\author[1]{Shivam Barwey*}
\author[1]{Malik Hassanaly}
\author[1]{Venkat Raman}
\author[2]{Adam Steinberg}
\affil[1]{Department of Aerospace Engineering, University of Michigan, Ann Arbor}
\affil[2]{Department of Aerospace Engineering, Georgia Institute of Technology, Atlanta}

\begin{document}
\maketitle

\begin{abstract} 
This work utilizes data-driven methods to morph a series of time-resolved experimental OH-PLIF images into corresponding three-component planar PIV fields in the closed domain of a premixed swirl combustor. The task is carried out with a fully convolutional network, which is a type of convolutional neural network (CNN) used in many applications in machine learning, alongside an existing experimental dataset which consists of simultaneous OH-PLIF and PIV measurements in both attached and detached flame regimes. Two types of models are compared: 1) a global CNN which is trained using images from the entire domain, and 2) a set of local CNNs, which are trained only on individual sections of the domain. The locally trained models show improvement in creating mappings in the detached regime over the global models. A comparison between model performance in attached and detached regimes shows that the CNNs are much more accurate across the board in creating velocity fields for attached flames. Inclusion of time history in the PLIF input resulted in small noticeable improvement on average, which could imply a greater physical role of instantaneous spatial correlations in the decoding process over temporal dependencies from the perspective of the CNN. Additionally, the performance of local models trained to produce mappings in one section of the domain is tested on other, unexplored sections of the domain. Interestingly, local CNN performance on unseen domain regions revealed the models' ability to utilize symmetry and antisymmetry in the velocity field. Ultimately, this work shows the powerful ability of the CNN to decode the three-dimensional PIV fields from input OH-PLIF images, providing a potential groundwork for a very useful tool for experimental configurations in which accessibility of forms of simultaneous measurements are limited.

\blfootnote{*Email: sbarwey@umich.edu}
\blfootnote{Preprint accepted to Combustion Science and Technology.}
\end{abstract}

\section{Introduction}
Gas turbine combustors demonstrate complex physics in the form of multiscale mechanisms, which are driven primarily by turbulence-chemistry interactions and thermoacoustic oscillations. These mechanisms induce highly chaotic dynamical behavior in the system, leading to severe (often untraceable) flame stability issues that can have detrimental structural impact on the gas turbine in practical applications even for the smallest of changes in operating conditions. In swirl-stabilized combustors, for example, the interaction of multiple recirculation zones with flame propagation significantly impacts the stabilization process. Two different stabilization mechanisms are feasible: a) a shear layer stabilized flame resulting in a flame attachment to some feature of the geometry, and b) lifted flames that are stabilized near the stagnation surface formed between inflowing gases and an inner recirculation zone \cite{qiang16,qiang17,qiang18}. The instabilities observed in each of these regimes are due to pressure fluctuations, leading to high amounts of heat release rate variations within the domain which in turn affect the flow field. To better understand the physical aspects of these instabilities, measurements which contain information on the coupling between flame shape and flow field are required. To this end, simultaneous laser-based diagnostic techniques become especially useful \cite{adrian1992}. For example, combining planar laser-induced fluorescence (PLIF) of measureable species (i.e. OH) and particle image velocimetry (PIV) allows one to visualize directly the flame front and its induced effects on a projected true velocity field. As such, countless experimental studies in combustion utilize these simultaneous laser diagnostic measurements; examples are provided in Refs.~\cite{caux-brisebois, boxx2015, temme2014, steinberg2008measurements}. Ultimately, the utilization of simultaneous measurements allows the practitioner to deduce the physical causes of instantaneous spatial (or spatio-temporal) correlations between these measurements, leading to a better understanding of the underlying complex turbulence-flame interactions. In a broader sense, much of the physical inductions and analysis based on the correlations implied by simultaneous measurements 1) facilitate combustor design, 2) validate/parametrize models used in numerical simulations, and more recently, 3) provide the foundation for data-driven predictive models \cite{ramanEmergingTrends}.


In light of the impacts that diagnostic based techniques can have on the field of combustion as a whole, it is important to also mention some practical limitations of the above mentioned types of experimental data for operating conditions of interest which prohibit fully comprehensive analyses. For example, achieving desirable accuracy and precision in velocity field measurements is difficult due to the inherent complexity of the required experimental setups (e.g. low-signal lasers, inconsistent particle seed distributions, etc.). These complexities often lead to either temporal or spatial gaps in the already discrete planar time-series trajectory. To overcome such issues, many data analysis efforts based on the field of image processing have been utilized in the past to reconstruct missing data given a previous "incomplete" set of data. Common data-driven reconstruction techniques are simply interpolation methods, e.g. Delaunay triangulation, which have been applied to good effect \cite{chew1989constrained, song1999}. Another common route is to use methods based on proper orthogonal decomposition (POD), a powerful tool used in post-processing, image compression and reduced order modeling which revolves around preserving the high-energy features (in the least-squares sense) of the dataset via projection onto an optimally derived basis \cite{berkoozPOD, sirovich1987, steinbergPOD}. In this approach, the data is recast into time-averaged spatial basis functions, where the coordinates of the original data in this basis provide information on the temporal development of the flow. A particular variant as explored in the works of Refs.~\cite{everson1995, willcox2006, saini2016} called "gappy" POD has found success in velocity field reconstruction efforts due to its utilization of spatiotemporal coherence to fill in the missing gaps of the data. 

Most implementations of the above mentioned techniques are purely interpolative, however, and do not use other types data to aid in the reconstruction (i.e. temporal gaps in PIV data are filled in using interpolation methods of the same PIV data). This can be problematic when considering limitations associated with simultaneously measured data (i.e. simultaneous OH-PLIF and PIV) -- situations may arise in which one type of data is either a) sampled at a higher rate, or b) captures a smaller section of the combustor domain. Although multiple data sources can be integrated into these interpolative techniques by concatenating the data, (i.e. simultaneous PLIF and PIV snapshot can be reformulated as a single vector with length equal to the sum of dimensions of the individual fields, and the data analysis techniques may be applied to this extended or augmented field), this type of incorporation of simultaneously measured fields does not take advantage of the inherent correlation between the fields in the reconstruction. In other words, the premise of obtaining physical conclusions from simultaneously measured fields lies in the fact that the correlations between these fields can be visualized -- there could be much value, then, in deriving reconstructions of one type of measurement directly from another (i.e. an extrapolative versus interpolative reconstruction framework). The current work is focused on addressing this problem, and expands the existing body of study in reconstruction methods by utilizing the spatial information contained in one form of measurement (in this case, planar OH-PLIF) to directly map to another form of measurement (three-component planar PIV) in the domain of a swirl-stabilized combustor. Similar concepts regarding the general recovery of velocity fields from scalar fields, which rely on an inversion of the Navier-Stokes equations, have been explored in the past in Refs.~\cite{dahm1, dahm2} for experimental and numerical settings. The primary focus of this work is related to an alternative formulation of the mapping itself, and potential implications of the technique in the context of experimental diagnostics are discussed. For example, in the case that the PLIF data captures a wider region of the domain than the PIV (as seen here), such a mapping tool would prove very useful for extending or interpolating velocity field data without the need for additional physical measurements. 

The OH-PLIF to PIV mapping problem is formulated in a nonlinear regression setting, i.e. the underlying assumption is that if a true transformation between the two fields does indeed exist, this transformation depends on nonlinear functions of linear projection operators. In this work, the regression is to be accomplished with a fully convolutional network, which is a type of convolutional neural network (CNN) that has gained popularity in the machine learning community in recent years \cite{lecun1995convolutional, lawrence1997, krizhevsky2012imagenet}. Briefly, the primary appeal of the CNN over a traditional artificial neural network (ANN) based model in data processing applications is twofold: 1) the CNN model obeys translational invariance in the medium-to-small length scales as implied by learned features in the input field, and 2) this translational invariance allows the CNN to have a massively reduced parameter space when compared to the ANN, leading to more computationally efficient and conservative (with regards to overfitting) implementations of regression tasks. Common CNN applications in machine learning involve mapping an image (or a pixel of an image for segmentation problems) to some categorical distribution given by a probability mass function (PMF), i.e. classification tasks. CNNs are also emerging powerful tools in the fluid dynamics community; they have been applied in combustion settings in the context of combustion instability prediction and deconvolution in large eddy simulation to estimate sub-grid scale reaction rates \cite{sarkar_isu, poinsot_cnn}. 

This work utilizes the CNN in a nonlinear regression framework to effectively morph a series of single-channel time-resolved experimental OH-PLIF images into corresponding three-channel planar PIV fields in the closed domain of a premixed swirl combustor. More specifically, two types of models are compared: 1) a global CNN which is trained using images from the entire domain, and 2) a set of local CNNs, which are trained only on individual sections of the domain. The models are trained on flame shapes in both the attached and detached flame configurations. The effect of time history in the input, as well as the extent to which a mapping trained on one particular region of the domain can be extended to other, unseen parts of the domain is also studied. 

The paper proceeds as follows. In Sec.~\ref{sec:data}, the experimental apparatus, data preprocessing steps, and organization of the data into training and testing sets is discussed. In Sec.~\ref{sec:methodology} and \ref{sec:architecture}, a summary of the CNN methodology and the architecture used in the study is provided, respectively. Then, in Sec.~\ref{sec:results}, the mapping results and implications are discussed using unseen data at both the same and entirely different operating conditions as the training data. Lastly, in Sec.~\ref{sec:conclusion}, concluding remarks and future directions are provided. 

\section{Experiments and Dataset}
\label{sec:data}
The data used for this analysis was acquired in the experiments of Ref.~\cite{qiang_new}, which contains a detailed description on the combustor geometry and equipment used for the relevant diagnostics. A brief summary of the experimental configuration is provided here, as are descriptions of pre-processing steps, training, and testing data.

\begin{figure}
    \begin{center}
    \includegraphics[width=0.6\columnwidth]{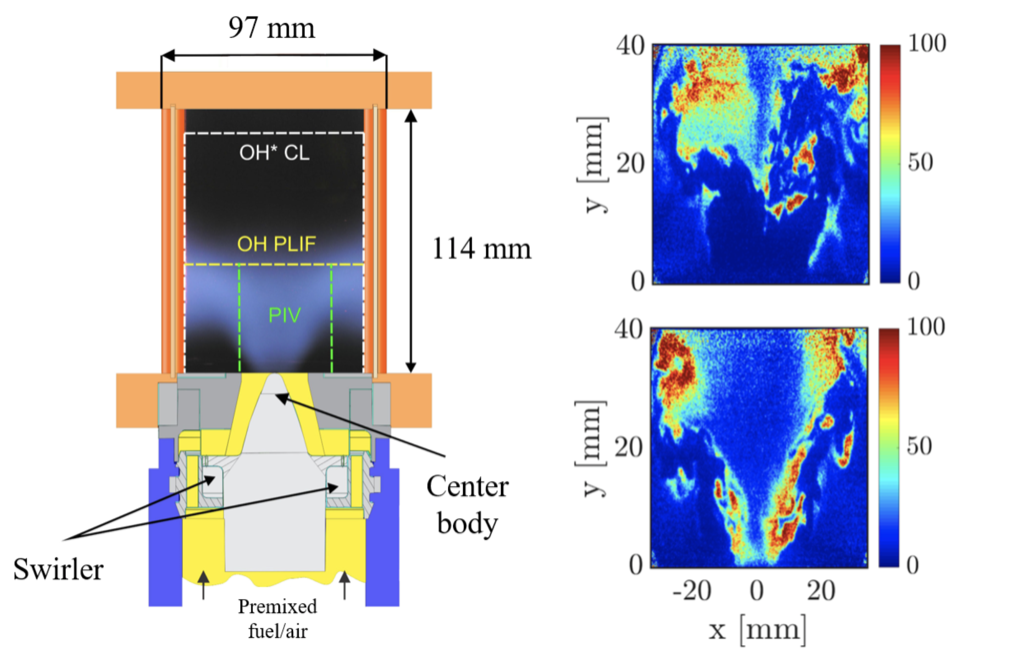}
    \caption{(Left) Schematic of combustor used to record the data. (Right) OH-PLIF examples of flame in detached (top) and attached (bottom) configurations.}
    \label{fig:schematic}
    \end{center}
\end{figure}

The gas turbine model combustor is shown in Fig.~\ref{fig:schematic}, which has been used in a number of experimental \cite{caux-brisebois, meier, oconnor1}, numerical \cite{koo_dlr}, and machine learning \cite{barwey_ctm} works to study the effects of thermoacoustic instabilities and flame transition. Premixed fuel and air are fed through a plenum to the radial swirler before entering the combustion chamber, where vortex breakdown generates a strong inner recirculation zone. The training dataset used in this study is characterized by a fuel-air mixture of equivalence ratio of $\phi = 0.60$, preheated to a temperature of 400~K. The fuel is made of 80\% CH$_4$ and 20\% CO$_2$ by volume. The air flow rate is 400~SLPM. This case was selected from the test matrix in Ref.~\cite{qiang_new} due its ergodic nature: for the same operating conditions (i.e. same equivalence ratio, mass flow rate), the flame shape observed spontaneous transitions between clearly defined attached and detached flame states.

Figure~\ref{fig:schematic} shows typical instantaneous OH PLIF images in attached (bottom) and detached states (top). In the attached state, high OH concentrations are present in the shear layers that separate the inner and outer recirculation zones, while in the detached state, a rotating helical vortex core generates a highly asymmetric OH field. The transitions are convenient in this neural network application, as training and testing can be done for a time-series that contains transitions between the attached and detached states at the same operating condition, allowing for a useful isolation of model performance in two different flame regimes. 


Data was collected using 10~kHz repetition-rate OH PLIF and S-PIV, providing simultaneous time-resolved 2D measurements of the OH radical distribution and of the three velocity components over a total measurement time span of 1.5~s, resulting in 15000 instantaneous snapshots for a single time series. It is apparent from Fig.~\ref{fig:schematic} that the OH-PLIF domain is larger and more symmetric than the PIV domain size. Furthermore, the pixel dimensions of the raw OH-PLIF and PIV images were originally $832 \times 503$ and $79 \times 53$ respectively. As a pre-processing step, each of the PLIF and PIV images were cropped to force their domains to be as close to each other as possible to allow for a well defined OH-PLIF to PIV mapping arrangement. As such, the resulting pixel dimensions of the cropped images were $472 \times 409$ for PLIF and $56 \times 51$ for PIV. 

\begin{figure}
    \begin{center}
    \includegraphics[width=0.8\columnwidth]{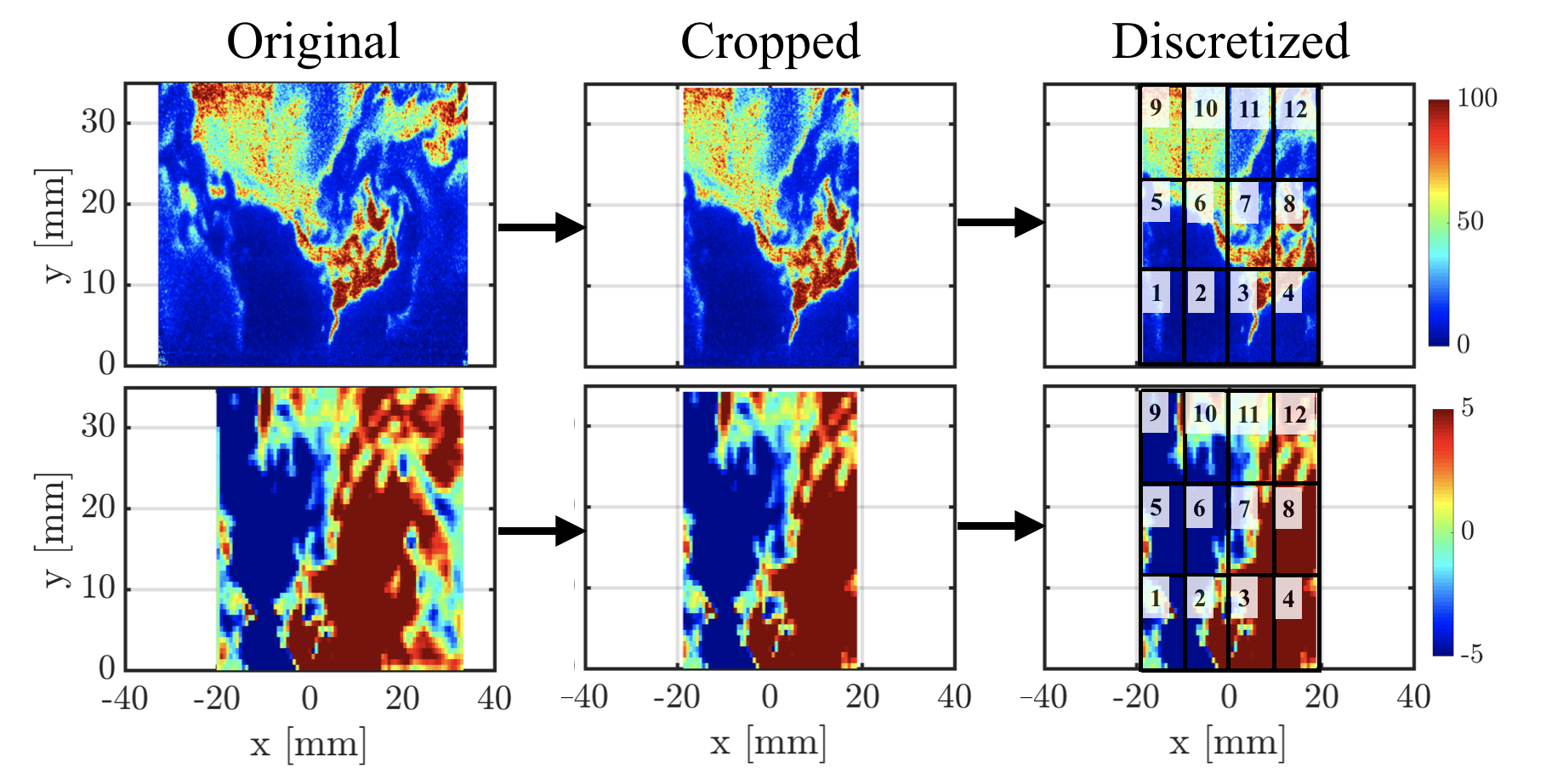}
    \caption{Original images and box discretizations. Upper row is PLIF, lower row is PIV (the same steps taken for PIV-x/y/z). OH in units of pixel intensity, velocity in m/s.}
    \label{fig:discretization}
    \end{center}
\end{figure}

Smaller sections resulting from a discretization of these cropped images into $N_b$ subsections, or boxes, are to be fed into the neural network for training and testing. This discretization step is shown in Fig.~\ref{fig:discretization}. Note that each "box" denotes a particular region of the original domain such that for each OH-PLIF box, a PIV box spanning roughly the same region in physical space is also available. This way, the CNN mappings between OH-PLIF and PIV fields can be easily analyzed as a function of domain location, allowing the model to conveniently isolate levels information contained in sub-samples of the domain. Furthermore, as will be seen in the results below, restricting the CNN operation to smaller subsets of the domain allows comparisons between local (box-specific) and global (box-agnostic) models. As shown in Fig.~\ref{fig:discretization}, the discretization is performed for $N_b = 12$ (with each box of equal size and nonintersecting), resulting in box sizes of  $118 \times 143$ and $14 \times 17$ for the OH-PLIF and PIV data respectively.



The full dataset of 15000 simultaneous OH PLIF and PIV images consists of roughly 8 transitions between attached and lifted states. The training set is a 14000-snapshot subset of this full time series that covers 7 of these 8 transitions. Furthermore, of the 14000 training images, roughly 8500 are in the detached state and 4000 are in the detached state. Note that the remaining 1500 snapshots are attributed to a non-instantaneous "transitioning" state which exists between the attached and detached states. The total number of available images after the discretization into $N_b=12$ boxes is equal to 168000 for each velocity component, since the discretization imposes 12 sub-images per timestep. 

Two testing sets are used to evaluate the accuracy of the model. The first testing set (referred to as testing set 1) is the remaining 1000-snapshot subset sourced from the time series as the training set described above. Therefore, in testing set 1, the combustor operating conditions are the same as those in the training set. Furthermore, a single attached-to-detached state transition is observed in these 1000 snapshots, where roughly 600 snapshots are attached flames, 300 are detached flames, and 100 are associated with the transitioning flame.

For a better evaluation of model robustness, a second testing set (testing set 2) is used at entirely different operating conditions from the training set. This data, which is also a 1.5 second time series sampled at 10kHz from the same combustor, is characterized by an equivalence ratio of $\phi = 0.65$, a preheat temperature of 300~K, fuel composition of 90\% CH$_4$ and 10\% CO$_2$ by volume, and air flow rate of 460~SLPM. Essentially, all major tunable operating conditions have been changed between the training set and testing set 2. This testing set also experiences a single spontaneous transition, and thus also produces both attached and detached flame shapes. More specifically, of the 15000 available images in testing set 2, there are roughly 9500 detached flames and 4500 attached flames (with about 1000 transitioning flames). 

Despite the fact that both testing sets include a single attached-to-detached flame state transition, the transition flame state itself will not be analyzed in the results below. In other words, only the subsets of attached and detached flames from each of the testing sets will be used to evaluate model performance.

\section{Methodology} 
\label{sec:methodology} 
In this section, the CNN methodology as it applies to this work is summarized. As the specifics of the CNN (especially in the fully convolutional context) are out of scope here, the reader is recommended to consult Refs.~\cite{dumoulin2016guide, goodfellow2016deep} for more detail on the subject.

A PLIF image associated with box $b$ (the model input) is denoted $X_b \in \mathbb{R}^{118 \times 143}$. A PIV image for a single velocity component from the same box (which is the desired model output) is denoted $Y_b \in \mathbb{R}^{14 \times 17}$. A fully convolutional neural network $\cal C$ is designed such that $\mathcal{C}: \mathbb{R}^{118 \times 143} \rightarrow \mathbb{R}^{14 \times 17}$. The elements of the parameter set $\Phi$ of $\cal C$ are to be estimated such that $\mathcal{C}(X_b | \Phi) = \tilde{Y_b} \approx Y_b$. 

The assembly of $\Phi$ is related directly to the \textit{hidden layers}, which are the main building blocks of any neural network. Each layer $L = 1,...,N_L$ performs the same operation,
\begin{equation}
    \textbf{y}_L = \beta (W_L \textbf{x}_L + \textbf{b}_L), 
    \label{eq:hidden}
\end{equation}
where $\textbf{y}_L$ is the vectorized output tensor, $\textbf{x}_L $ is the vectorized input tensor, $W_L$ is a matrix of weights, $\textbf{b}_L $ is a vector of biases, and $\beta$ is an activation function that acts element-wise (usually nonlinear). The collection of all weights and biases in all $N_L$ layers are the elements in $\Phi$. Generally, the size of $\Phi$ increases when either 1) row rank of $W_L$ (number of neurons) in a layer increases, or 2) the number of layers $N_L$ increases. The CNN input $X_b$ is propagated through a series of the $N_L$ hidden layers to eventually produce the output $Y_b$.

With the regression task in mind, to obtain an optimal estimate of $\Phi$ the model assumes that given $\cal C$, each pixel of the desired output tensor $Y_{b,i}$, where $i = 1, ..., N_p = 14*17$, is a random variable with the distribution
\begin{equation}
    Y_{b,i} | X_b, \Phi \sim \textnormal{Normal}(\mathcal{C}_i(X_b | \Phi), \sigma^2).
    \label{eq:likelihood}
\end{equation}
This generative distribution from which $Y_b$ is sampled is termed the data \textit{likelihood}. Given this Gaussian likelihood assumption, it is clear that the CNN prediction is a point-wise mean conditioned on the input data and the parameters of the network. The scalar $\sigma$ is the standard deviation, which can be interpreted as an epistemic uncertainty associated with each pixel measurement in the training data -- in practical applications, its exact value is unknown. 
If the training data are independent and identically distributed, it is straightforward to show that the log-likelihood of the full output tensor can be written as a quantity proportional to the mean-squared error (MSE) of the output tensor,
\begin{equation}
    \log p(Y_b|X_b, \Phi) \propto \frac{1}{N_p}\sum_{i=1}^{N_p} (Y_{b,i} - \mathcal{C}_i(X_b | \Phi))^2 \equiv \textnormal{MSE}.
\end{equation}
The optimal parameters $\Phi$ are determined through maximum likelihood estimation (MLE), which provides a point estimate of the likelihood mean. The MLE is performed by maximizing the log-likelihood (equivalent to minimizing the MSE in the case of a Gaussian likelihood) with respect to each element in $\Phi$ using gradient-based optimization methods. 

The above methodology is quite generic and is not just unique to CNNs. The key differences provided by CNNs when compared to traditional ANNs are clear when looking at the structure of the hidden layer weights $W_L$. In a CNN, the hidden layers are called \textit{convolutional layers}. In a convolutional layer, the matrix-vector operation $W_L \textbf{x}_L$ represents a discrete convolution of the input with a set of kernels. Each convolutional layer has its own set of kernels, and the weights in $W_L$ are contained in the kernels. The convolution operation is given by a sliding, channel-wise dot product (which is implemented as a cross-correlation) of the scalars in the kernel with the corresponding scalars in the overlapping section of the input tensor for the whole domain.

The advantage of the CNN comes in this channel-wise convolution operation and its so-called parameter-sharing property (the weights in kernel are \textit{shared} across all input channels). The convolutional layer observes translational invariance at the lengthscales implied by the filter sizes -- the output identifies features of the input present in all of its channels regardless of where these features occur spatially in the input. Note that each kernel in a layer corresponds to one feature of the input in the same layer. Because the kernels are expected to identify features in some location of the input, they are constrained to sizes much smaller than the input but are by design larger than (or as large as) the identified feature itself. Such behavior is useful in producing the desired PLIF $\rightarrow$ PIV decoder, as it provides 1) increased sparsity in $W_L$, and 2) less \textit{unique} weights in the already sparse $W_L$. Instead, if one were to use a conventional ANN, the weight matrix would be fully populated and each weight would be unique, which could possibly lead to severe overfitting as a consequence of overparameterization and/or immense computational expense due to increased training times / training dataset sizes as a result of the larger parameter space. Further, the fully populated $W_L$ means that in a trained ANN, an identified feature of the input would not be recognized as the same feature if it occurred in a different spatial location, which could be a restriction in the PIV regression process.

\section{Network Architecture}
\label{sec:architecture}
A widely used practice in deep CNN architectures is to progressively downsample the input through a series of convolutional layers, often to the point at which a the tensor width and height of a final resulting output is on the order of 1x1. This downsampling process is the called the encoding phase. As this work is interested in producing 2D image output from a 2D image input, some form of upsampling is also eventually required. This upsampling is referred to as the decoding phase. The combination of the encoding and decoding phases constructs the fully convolutional network, i.e. the CNN given by $\cal C$. 

Decoding in CNN-based architectures can be achieved via upsampling in many ways. Learnable upsampling with transpose convolution layers, which applies a similar sparse and parameter-sharing weight matrix structure as a normal convolutional layer, is used here. These layers can also be generically described by Eq.~\ref{eq:hidden} -- the weight matrix, however, is designed such that the output vectorized tensor is of higher dimension than the input. The details of the transpose convolutional kernels and their implementation are not discussed here, but can be found in Ref.~\cite{dumoulin2016guide}. The main takeaway is that a layer which incorporates transpose convolutions provides an upsampling method which preserves the sparsity and parameter-sharing properties of a normal convolutional layer.

\begin{figure}
    \centering
    \includegraphics[width = \columnwidth]{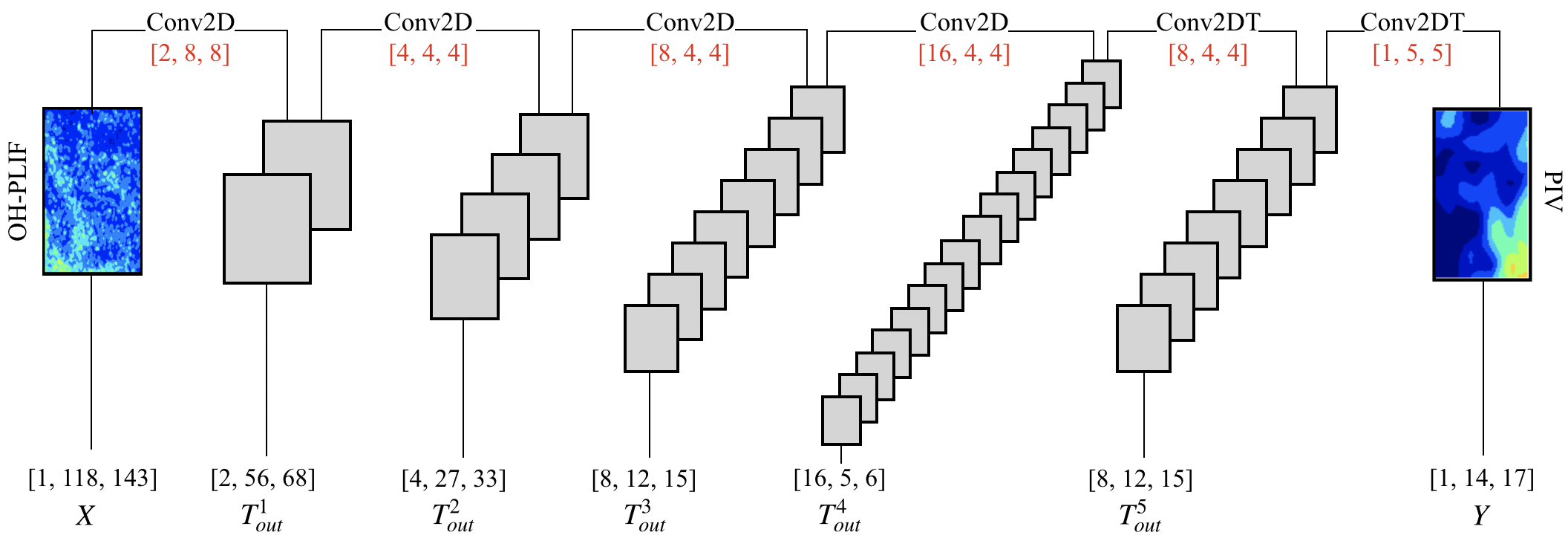}
    \caption{Fully convolutional network architecture. Kernel tensor shapes for each layer are given in red, below the layer title. Tensor shapes for each layer are displayed in the lower row: for example, $T_{out}^1$ is the output tensor of layer 1 and input tensor of layer 2. Conv2D refers to convolutional layer, and Conv2DT refers to a transpose convolutional layer.}
    \label{fig:network}
\end{figure}

As such, the architecture of the CNN $\cal C$ used here is shown in Fig.~\ref{fig:network}. The rectified linear unit (ReLU) nonlinearity is used for $\beta$ after each convolutional and transpose convolutional layer, except for the last layer where a linear $\beta$ is used. Kernel tensors for each layer are square, and their shapes are specified in red text in Fig.~\ref{fig:network}. Note that the network outputs a single-channel PIV tensor $Y_b$, which means that different CNNs are trained for different velocity components. 

The gradient of the MSE loss function with respect to the parameters is found using backpropagation – its mathematical details can be found in Ref.~\cite{hecht1992theory}. For parameter updates, the Adam optimizer was used with a learning rate of $0.001$ \cite{kingma2014adam}. Initial guesses for parameters were obtained with Xavier initialization. Training samples were randomly shuffled with 10\% of the training data set aside for validation to monitor overfitting. The neural networks were implemented with the open-source library PyTorch \cite{paszke2017automatic}.

For the same network architecture in Fig.~\ref{fig:network}, two types of CNNs are compared in the results below:
\begin{enumerate}
    \item A "global" CNN, which is trained using images from all twelve boxes. Total number of training images is 168000, as described in Sec.~\ref{sec:data}. 
    \item A "local" CNN, which is a function of an individual box. For a given box, the local CNN is trained only using images for that particular box. Due to this restriction, the total number of training images per local CNN is 14000. 
\end{enumerate}

Since each velocity component is treated independently, the results below come from a total of 3 global CNNs (one for each velocity component) and 36 local CNNs (one for each of the twelve boxes with three velocity components per box). 

\section{Results}
\label{sec:results}
The training and testing results are described here for both global and local CNN implementations. Specifically, decoding results and accuracy from testing set 1 (same operating conditions) and testing set 2 (different operating conditions) are discussed within the global and local model framework. Testing results for both flame shape regimes are compared, and the improvements gained from the use of local CNNs are also highlighted. The effects of time history inclusion on the local models then follows. Lastly, the ability of local CNN models to extrapolate to other, unexplored domain regions is shown and analyzed. An interpretation of the PIV decoding from the perspective of proper orthogonal decomposition is provided in the Appendix.

\subsection{Same Operating Conditions}
\label{sec:results_set1}
\begin{figure}
    \begin{center}
    \includegraphics[width=\columnwidth]{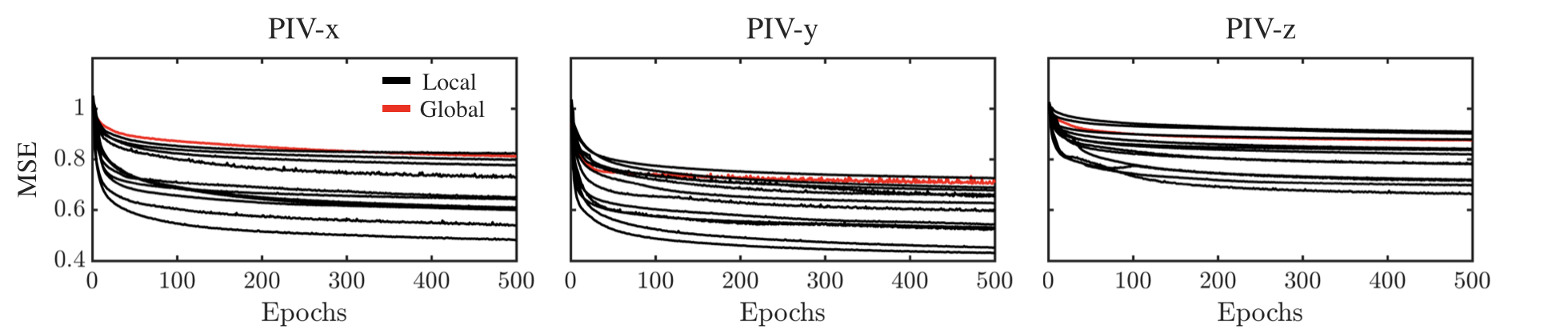}
    \caption{Training results for global and local CNNs for the three velocity components. $\mathcal{C}_{G,x}$, $\mathcal{C}_{G,y}$, and $\mathcal{C}_{G,z}$ are indicated by red lines, and black lines indicate the local models.}
    \label{fig:training_mse}
    \end{center}
\end{figure}

The training results for the global and local CNNs trained for each PIV component are shown in Fig.~\ref{fig:training_mse}. In this context, the MSE is obtained by averaging squared errors both over all pixels each input image and all input images in the training set. Each epoch on the x-axis corresponds to a single pass through all training set images, and the MSE loss function as a result of this pass is computed. The intent of Fig.~\ref{fig:training_mse} is to show a convergence of the parameters of the CNN (convolution kernels) to a local minimum in the parameter space. Since the loss functions appear to be well-converged for each of global and local models, it is safe to assume that the models have been sufficiently optimized in the context of the training data. Note that MSE value here are unitless as the model inputs have been normalized to zero mean and unit standard deviation. 

\begin{figure}
    \centering
    \includegraphics[width = \columnwidth]{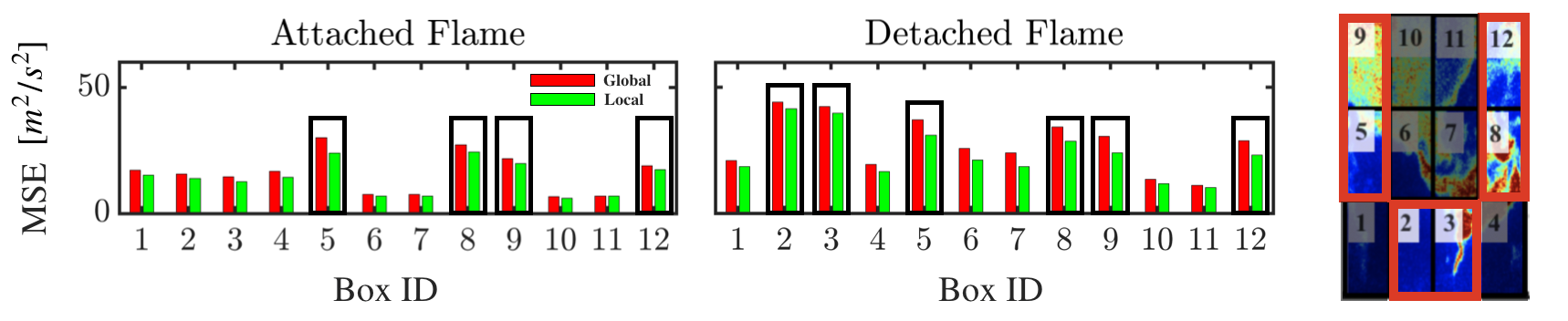}
    \caption{MSE values for test set 1 in attached (left) and detached (right) regimes for global (red) and local (green) CNNs. These MSE values are unscaled and are averaged over all three velocity components. Shown on the far right is the box discretization from Fig.~\ref{fig:discretization} for convenience. High-error boxes discussed in text are enclosed.}
    \label{fig:set1_mse}
\end{figure}

For both global and local models, the MSE evaluated for testing set 1 is shown in Fig.~\ref{fig:set1_mse}. Recall that the source time series used to obtain testing set 1 is at the same operating conditions used in the training data. For a macroscopic assessment of local versus global performance on unseen data at the same operating conditions, the MSE values in Fig.~\ref{fig:set1_mse} are averaged across all pixels, snapshots and velocity components in the training data. Further, the MSE values are unscaled (units in $m^2/s^2$). Note also that the MSE values are computed as a function of box number, which is essentially an indicator as to how each model performs for a particular region in the domain. 

A quick glance at Fig.~\ref{fig:set1_mse} shows two major trends: 1) the local model outperforms the global model in terms of MSE averaged over velocity components for the same operating conditions, and 2) component-averaged errors in the detached regime are generally higher across the board when compared to attached. This implies that localizing the model parameters (i.e. kernel weights and biases) to specific regions of the flow domain can result in better velocity field decodings from the PLIF images for unseen data at the same operating conditions. It also implies that the decodings perform worse overall in the detached regime, which is understandable, as the detached flame dynamics span a much larger subset of the PLIF/PIV phase space. Note that the boxes used to discretize the domain in Sec.~\ref{sec:data} were chosen naively -- there may exist more refined forms of nonintersecting discretizations (not-necessarily of equal size) that can result in much better localized model performance in the detached regime. 

A more detailed inspection of Fig.~\ref{fig:set1_mse} shows noticeably higher errors in boxes 5, 8, 9, and 12 in the attached regime than in the other boxes. In the detached regime, it is these same boxes that also appear with high error along with two additional boxes: 2 and 3. These boxes are indicated in the enclosed red regions in Fig.~\ref{fig:set1_mse}. It is expected that these high-error boxes in both attached and detached regimes observe a symmetry in the combustor domain, since the attached flame is largely symmetric and the detached flame observe periodic dynamics about the x=0 axis. In the attached regime, for instance, many of the turbulence-chemistry fluctuations in the flame are expected to occur in the outer regions of the V-shaped flame, and this region is indeed enclosed by boxes 5, 8, 9 and 12. Furthermore, the periodic asymmetric quality of the detached flame dynamics is captured mostly in these same four boxes. The two additional high-error boxes of 2 and 3 in the detached regime correspond directly to the burner exit; these same boxes have expectedly much lower errors in the attached regime. The high errors at the burner exit in the detached flame are a direct result of the absence of significant OH-PLIF signal in the presence of a highly unsteady flame anchoring point driven by the precessing vortex core. In other words, the OH correlations at the burner exit in the detached regime cannot translate to efficient velocity field encodings, as the dynamical variation in the portions of boxes 2 and 3 closest to the burner exit in the velocity field is much greater than that of OH in the same region.

In the attached regime, boxes 6, 7, 10, and 11 are lowest in error as indicated by Fig.~\ref{fig:set1_mse}. It is in these boxes where the "inside" of the V-shaped flame is captured (within the inner shear layer), leading to minimal variations in velocity field. In the detached regime, errors associated with boxes 6 and 7 increase significantly where errors in boxes 10 and 11 remain low. The increase in box 6 and 7 error is associated with the expected complexity in this region in the event of extreme flame asymmetry (in particular, these boxes would pick up any resultant asymmetry in the inner shear layer), a phenomenon which is much more prominent in the detached regime.

It should be noted that assessing model performance purely based on component-averaged MSE values does not give a complete picture of robustness. More insight into model performance can be gained with a qualitative analysis of 1) the velocity field reconstructions for a single given input PLIF image and 2) the time-averaged error fields plotted over the domain.

\begin{figure}
    \centering
    \includegraphics[width = \columnwidth]{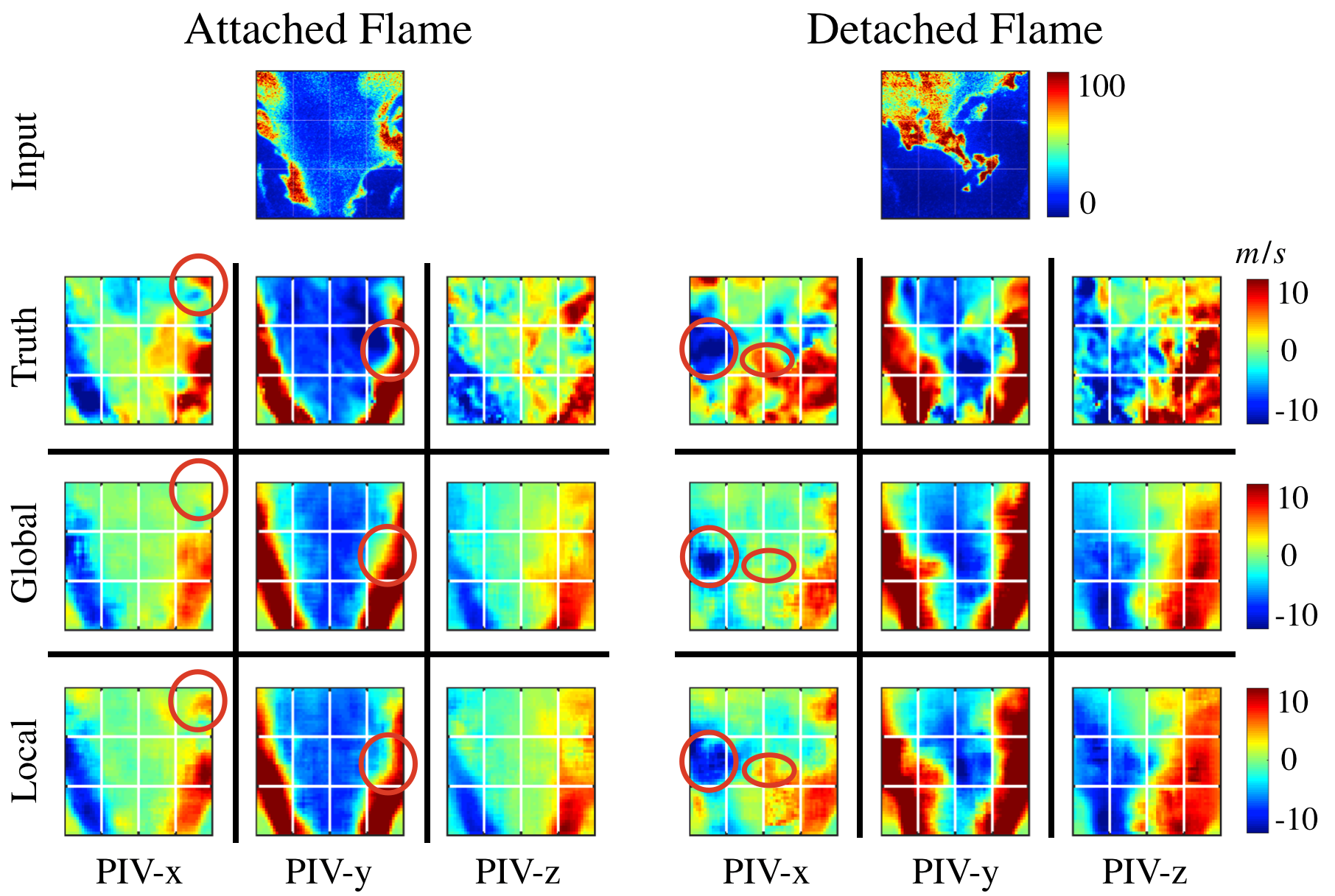}
    \caption{(Top) Sample OH-PLIF input from testing set. (Bottom) Corresponding PIV reconstructions for all three components using the global and local CNNs.}
    \label{fig:testing_1_inst}
\end{figure}

The velocity field decodings for a single OH field are shown in Fig.~\ref{fig:testing_1_inst}. The topmost row in Fig.~\ref{fig:testing_1_inst} displays a sample PLIF image for both attached (left) and detached (right) configurations with white lines delineating the box discretization cutoffs, and the lower rows compare corresponding velocity field reconstructions using both global and local CNN models to the ground truth velocity field (with same box delineations) for all three PIV components.  

In the attached flame velocity field reconstructions (left of Fig.~\ref{fig:testing_1_inst}), the improvements of the local CNN decodings implied by Fig.~\ref{fig:set1_mse} can be seen -- for the given PLIF input, local CNNs provide a slightly better representation of the true PIV fields for the PIV-x and y components. Indications of specific improvements observed in the local model are given by the red circles in Fig.~\ref{fig:testing_1_inst}. Specifically, the fluctuations in the top-most region of the domain are better resolved by the local model for x-component. In the y-component, small curvatures of velocity at the inner shear layer are captured by the local model that are not captured by the global model. However, in the z-direction, no noticeable difference in the attached regime is present between global and local models. In fact, in the attached regime, despite the above indicated differences, the global CNN does an adequate job in reproducing the relevant large-scale features such as the structure of positive/negative velocity streaks in the lower regions of the domain in PIV-x and z, as well as the general structure of the inner shear layer as implied by PIV-y. For PIV-z in particular, reconstructions near the burner exit are quite poor for both models. 

In the detached flame decodings on the right of Fig.~\ref{fig:testing_1_inst}, improvements provided by the local model are present especially in the PIV-x reconstructions; for example, the representation of the positive and negative velocity patterns better match the ground truth in the local model for PIV-x (indicated by red circles in Fig.~\ref{fig:testing_1_inst}). In PIV-y and PIV-z, global and local CNN predictions of the velocity fields are quite similar. Interestingly, the models are able to produce quite well the large-scale patterns of the asymmetric velocity fields, such as the S-shaped structure of the inner shear layer in PIV-y. However, many of the velocity field fluctuations (especially in the top most regions of the domain) are filtered out by the CNNs. This filtering effect is especially apparent in the PIV-z decodings. Furthermore, complexities near the burner exit (especially in PIV-x and PIV-z) are not captured well. Despite this, both the local and global CNNs show good promise in capturing the large-scale patterns and macroscopic structures associated with velocity fields in the detached regime. 

\begin{figure}
    \centering
    \includegraphics[width = \columnwidth]{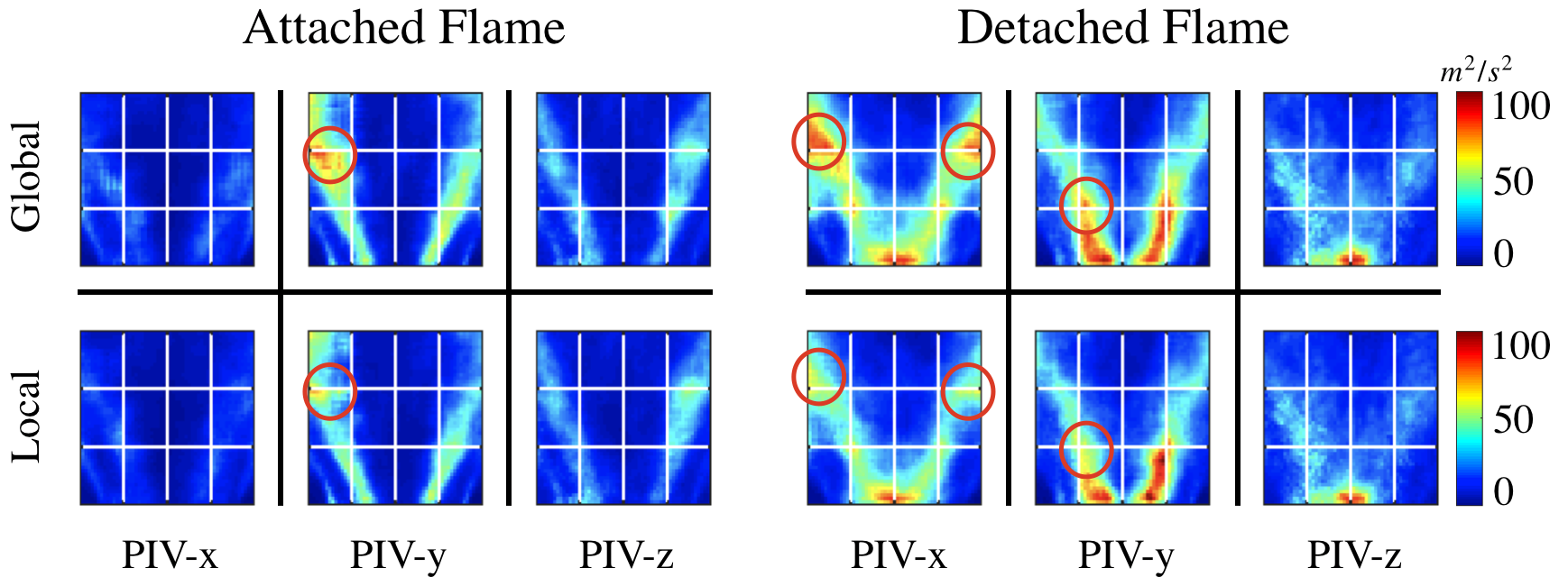}
    \caption{Time-averaged squared errors for attached (left) and detached (right) regimes in testing set 1. Top row is global model, bottom is local.}
    \label{fig:testing_1_timeavg}
\end{figure}

To better visualize model performance on the entire testing set, the time-averaged mean squared error fields for the testing set 1 time series in the attached and detached regimes are shown in Fig.~\ref{fig:testing_1_timeavg}. The lower errors in the attached regime are clearly apparent here. Figure~\ref{fig:testing_1_timeavg} reveals especially lower errors in the PIV-x reconstructions in the attached regime. The improvement of the local model in the attached regime across testing set 1 can be seen clearly in the PIV-y decodings, where there is a lower concentration of error in the box 5 and box 6 region (circled in Fig.~\ref{fig:testing_1_timeavg}). Similar to the attached flame case, the improvement in the local CNN framework for the detached flame is seen only in PIV-x and PIV-y, where the error fields for the PIV-z decodings are quite similar between local and global CNNs (notable regions circled in red). In particular, the errors in boxes 5, 8, 9 and 12 in the PIV-x and PIV-y case are dampened by the local CNN implementation. Lastly, the recurring theme of high error concentrations near the burner exit in the detached flame decodings (bottom-left and bottom-right corners of boxes 2 and 3, respectively) is again apparent in both local and global settings.



It should be noted that the local CNN models are indirectly overfitting to a particular region of the domain, a consequence intrinsic to their training procedure. The global CNNs are by design insensitive to domain location in the input. The trade-off presented by the local CNNs, assuming one is operating in a fixed domain with bounds known a-priori, is in overall reconstruction accuracy (see Fig.~\ref{fig:training_mse}). Any improvements seen in the local CNN output imply that patterns observed in the input OH field that lead to the velocity field decoding are more domain-dependent. Conversely, since little improvement is seen in the local CNNs over the global CNN in the PIV-z decoding, the identified OH features relevant for the PIV-z mapping are much less dependent on spatial domain location (i.e. these learned features are not a function of box number). It is possible that the global CNN can better compensate for its higher errors in PIV-x and PIV-y decodings with an increase in the complexity of the network architecture, though this would increase the required number of training epochs to reach convergence to optimal model parameters (which would in turn increase required training times). Lastly, it is interesting that both the models perform an acceptable job at reconstructing the full PIV images in the 12 boxes without large amounts of discontinuity between the boxes, despite the fact that the velocity field continuity between boxes was not explicitly enforced during training. 

\subsection{Different Operating Conditions}
\label{sec:results_set2}
In this section, similarly structured results as in Sec.~\ref{sec:results_set1} will be shown for testing set 2, the dataset which was recorded at entirely different operating conditions from the training set (see Sec.~\ref{sec:data} for dataset details). Velocity field decodings for this second dataset will be illustrated again from a local versus global CNN perspective, for both attached and detached flame configurations. 

\begin{figure}
    \centering
    \includegraphics[width = \columnwidth]{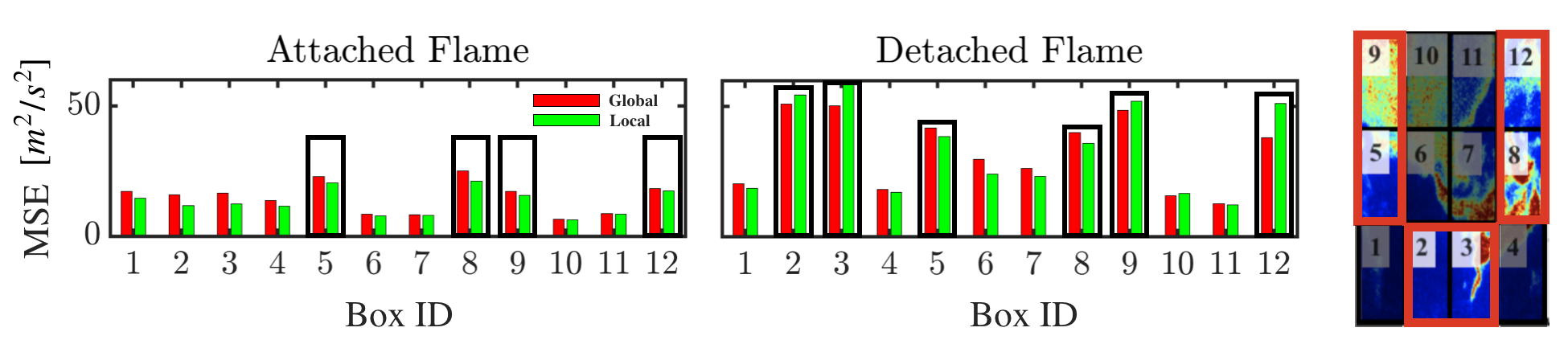}
    \caption{MSE values for test set 2 in attached (left) and detached (right) regimes for global (red) and local (green) CNNs. These MSE values are unscaled and are averaged over all three velocity components. Shown on the far right is the box discretization from Fig.~\ref{fig:discretization} for convenience. For consistency, the same boxes highlighted in this figure are the same as in Fig.~\ref{fig:set1_mse}}
    \label{fig:set2_mse}
\end{figure}

The mean squared errors of testing set 2 averaged over all three velocity components in both detached and attached configurations is shown in Fig.~\ref{fig:set2_mse}. This can be directly compared with the same values derived from testing set 1 (same operating conditions) shown in Fig.~\ref{fig:set1_mse}. Interestingly, the MSE values associated with boxes 5, 8, 9 and 12 in the attached regime have in fact \textit{decreased} in value from the Fig.~\ref{fig:set1_mse} counterparts in both global and local models. This is very encouraging, as one would expect the trained model to perform worse in any case on unseen data at different operating conditions. On the other hand, in the detached flame case, the errors have in fact increased over the Fig.~\ref{fig:set1_mse} counterparts. Most notably, the local model performs worse than the global model for testing set 2 in boxes 2, 3, 9, and 12. 

\begin{figure}
    \centering
    \includegraphics[width = \columnwidth]{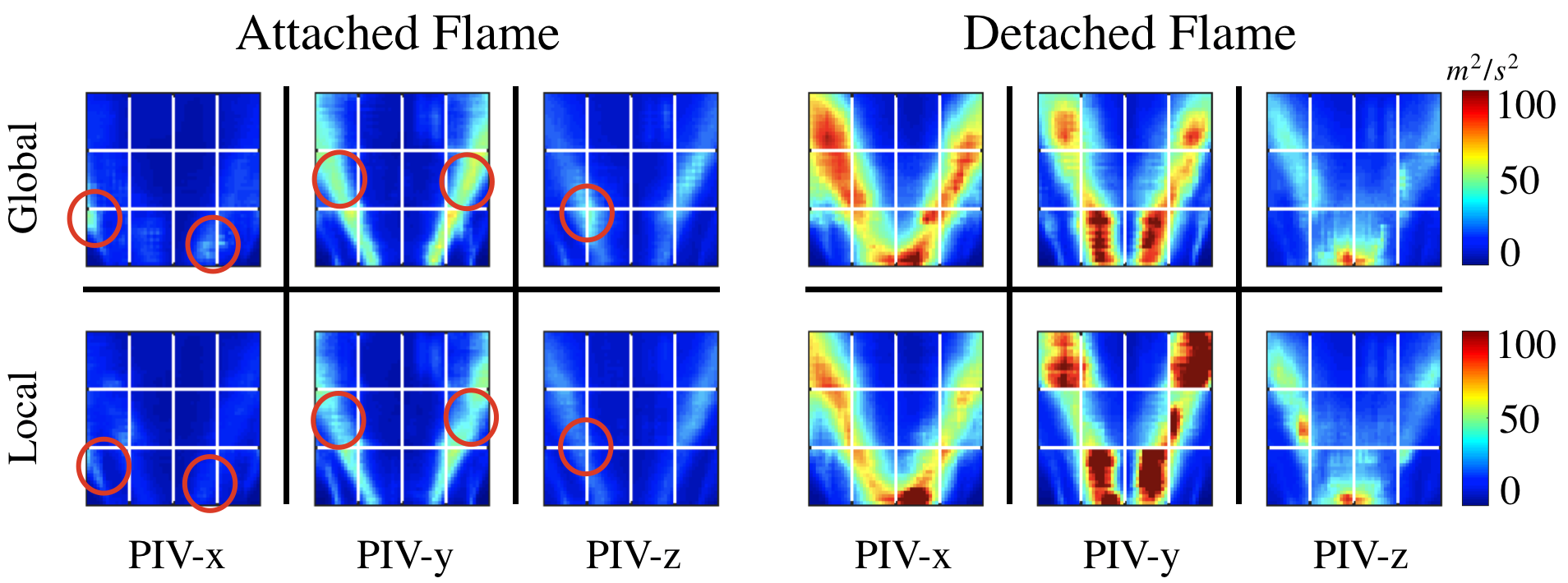}
    \caption{Time-averaged squared errors for attached (left) and detached (right) regimes in testing set 2. Top row is global model, bottom is local.}
    \label{fig:testing_2_timeavg}
\end{figure}

The time-averaged squared error fields for testing set 2 are shown in Fig.~\ref{fig:testing_2_timeavg}, which is analogous to Fig.~\ref{fig:testing_1_timeavg} for testing set 1. Notable regions of error reduction in the attached regime provided by the local model on this dataset are enclosed in red circles. An important point is that the large concentration of error present in testing set 1 for the PIV-y decoding in the attached flame (near boxes 5 and 8, highlighted in Fig.~\ref{fig:testing_1_timeavg}) is not present in Fig.~\ref{fig:testing_2_timeavg}, which affirms the conclusions made above regarding an improvement in the attached flame predictions of velocity field for unseen data at different operating conditions. It is encouraging that the CNN is able to strongly decode velocity fields in this setting at an even higher accuracy in some cases for data at different operating conditions, at least in the attached flame configuration. This strengthens the universality of the model. Errors in the detached flame are much larger for PIV-x and PIV-y predictions when compared to the testing set 1 counterparts. PIV-z error fields in Fig.~\ref{fig:testing_2_timeavg} are surprisingly quite similar to those found in Fig.~\ref{fig:testing_1_inst} -- most of the error contribution in the detached regime in Fig.~\ref{fig:set2_mse} stems from the PIV-x and y components. The local CNN is considerably worse in boxes 2, 3, 9, and 12 in the PIV-y predictions, which indicates an amount of overfitting of the local models to the training dataset.

Interestingly, the results from Fig.~\ref{fig:set2_mse} and \ref{fig:testing_2_timeavg} imply that the detached regime at different operating conditions utilizes different physical structures in the input OH field to produce the desired velocity field outputs. In other words, the trained convolution kernels are inaccurate in capturing the relevant physical features of the input PLIF image to produce the desired velocity fields in the detached regime at different operating conditions. On the other hand, in the attached regime, the relevant features in the OH field learned in the training phase to produce the velocity field mappings can be successfully extrapolated to different operating conditions. This means that the CNN is more "universal" when applied to attached flames, which could be due to the high level of dynamical steadiness in the attached configuration when compared to detached. 

\subsection{Inclusion of Time History}
\label{sec:results_timehist}
For the results above, the CNN architecture (shown in Fig.~\ref{fig:network}) takes in as input a single OH-PLIF image at the same time step as the desired velocity field output. It is reasonable to expect that accounting for memory in the OH field could improve the velocity field mapping results. This idea is explored in this section. Here, the network architecture is modified to allow for the inclusion of time history in the PLIF input to obtain the desired PIV output. Time history is accounted for by increasing the number of channels in the input PLIF image, where an additional channel would represent an OH field at some previous time step. To better isolate the effects of time history, the only modification to the network architecture is the input channel number (there is no addition of convolution kernels to the CNN, so the size of the model parameter space is unmodified).

For simplicity, the results below show the effects of adding only a single previous time-step to the PLIF input. More specifically, for an existing PLIF input measured at time $t$, an additional feature is added to the input PLIF field at time $t - dt$, where the desired output is the corresponding velocity field at time $t$. For conciseness, since the findings apply for both global and local models, the time history modification will be shown only for the local CNN framework. As such, the training results of the local CNNs with time history included in the input will be compared to the previous local CNNs without time history. The same two networks will then be compared with both testing sets 1 and 2 in terms of velocity field predictions, as done in Sec.~\ref{sec:results_set1} and \ref{sec:results_set2}. 

\begin{figure}
    \centering
    \includegraphics[width = \columnwidth]{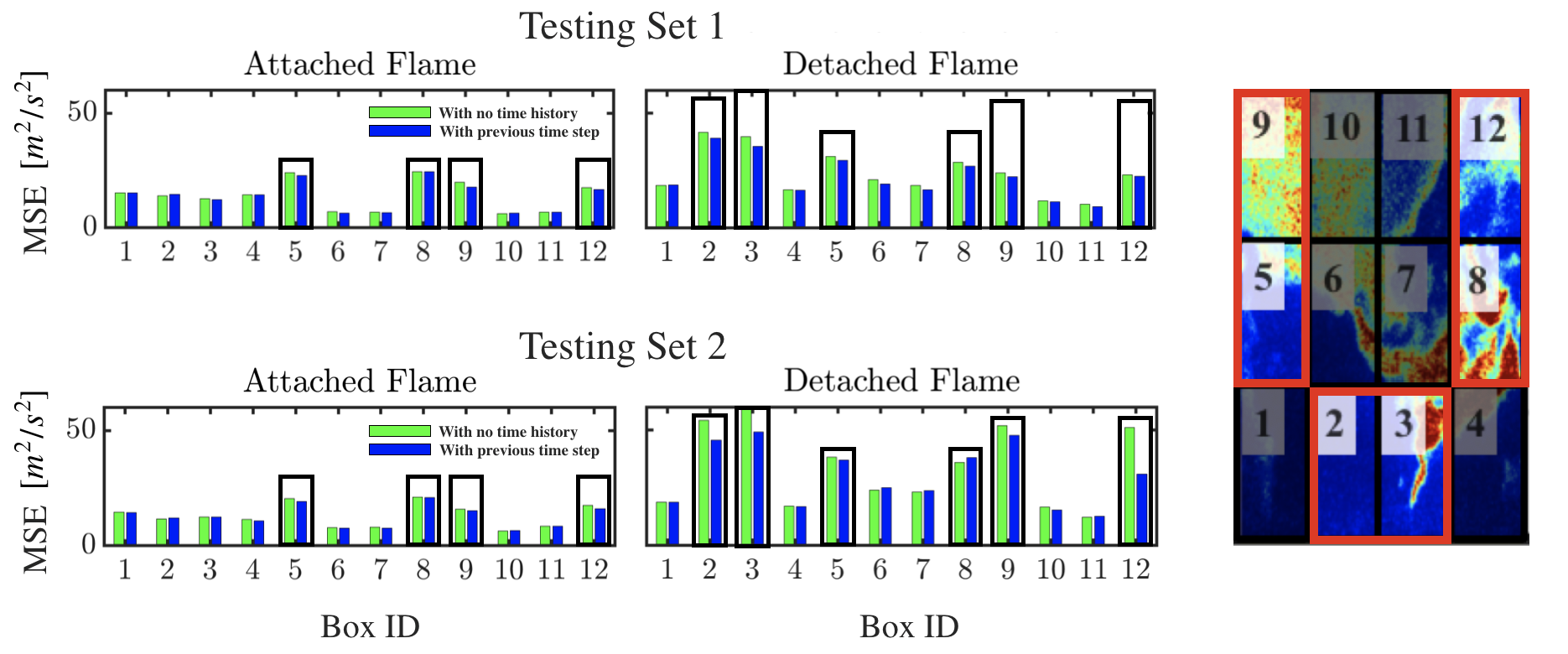}
    \caption{MSE values averaged over all three velocity components for testing set 1 (upper row) and set 2 (lower row). Green bars indicate local CNN results without time history, blue bars indicate new local CNN results with time history of one previous time step. Boxes of interest indicated in same manner as Figs.~\ref{fig:set1_mse} and \ref{fig:set2_mse}.}
    \label{fig:timehist_mse}
\end{figure}

Analogous to Figs.~\ref{fig:set1_mse} and \ref{fig:set2_mse}, the MSE values averaged over the three velocity field components for the local CNN with (blue bars) and without (green bars) time history are shown in Fig.~\ref{fig:timehist_mse}. In testing set 1 (same operating conditions), it can be seen that the local CNN model with a single previous time step of history included does provide a slight improvement in the attached and detached regimes, though not by much. A similar trend carries over into testing set 2 (different operating conditions), where again there is minimal accuracy gain due to time history. The only exception is in box 12 for the detached regime in testing set 2, where significant improvement is observed.  

\begin{figure}
    \centering
    \includegraphics[width = \columnwidth]{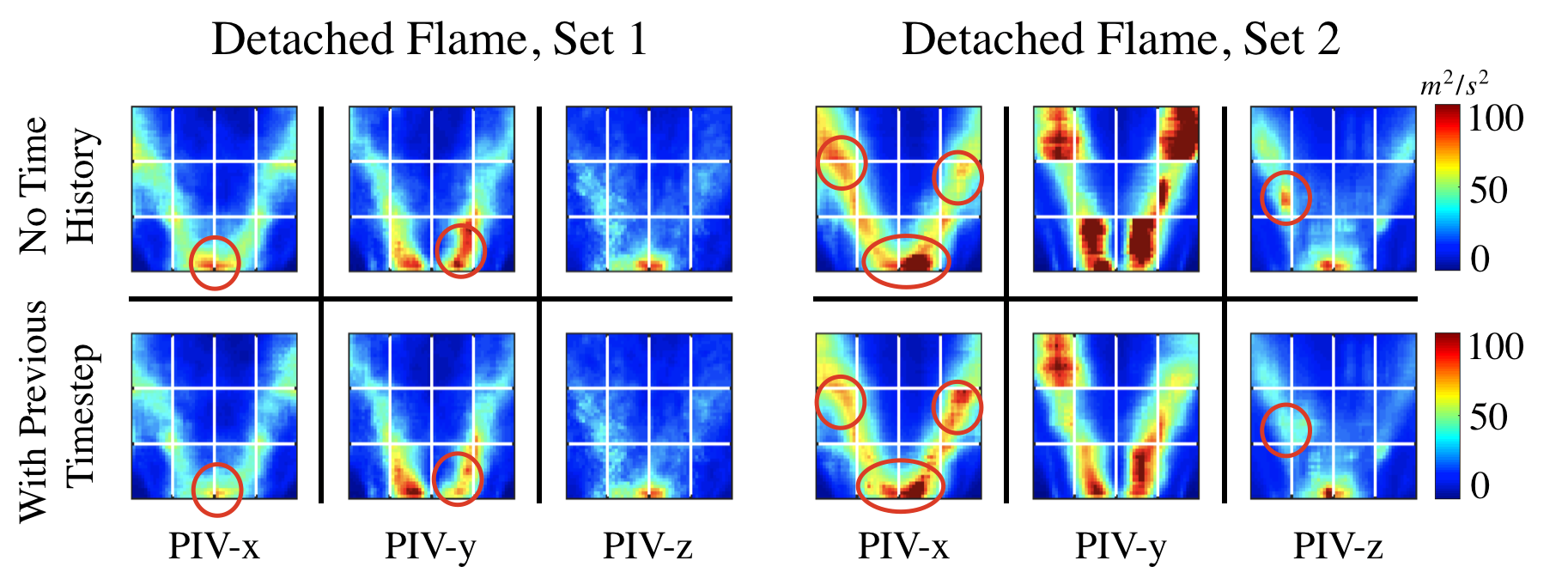}
    \caption{Time-averaged squared errors for testing set 1 (left) and set 2 (right) in the detached flame regime. Top row is local CNN error without time history, bottom row is with previous OH field timestep included.}
    \label{fig:timehist_timeavg}
\end{figure}

For a closer look at the error distribution, the time-averaged error fields for both testing sets are shown in Fig.~\ref{fig:timehist_timeavg} for the detached regime only. Again, minimal improvements are seen (except in the case of PIV-y Box 12 for testing set 2, which could be due to issues in the parameter convergence of the local model). Regions of interest are indicated via red circles. For testing set 1 (left of Fig.~\ref{fig:timehist_timeavg}), the slight improvements near the burner exit are observed, and little to no visible improvement in the PIV-z decodings in a time-averaged sense. For set 2, time history in the OH field input appears to improves predictions in the PIV-y field. Furthermore, unlike testing set 1, testing set 2 displays slight improvement provided by the time history in the z-direction (namely box 5, circled in red), though this is not as significant. Small improvements are also seen in PIV-x near the burner exit and in boxes 5/9; the model with time history leads to more error, however, in box 8 -- this could be related to issues in convergence of the box 8 model parameters. 

Ultimately, it is difficult to assess directly whether the time history inclusion is truly effective (the only clear case for improvement is seen in PIV-y for set 2). One would expect time history information, at the very least, to improve detached flame decodings across the board (for both sets 1 and 2), as the time history contains more information on the induced effects on the velocity field of the periodic dynamics attributed to the helical vortex core. The fact that this is not the case leads to an interesting interpretation of the decoding process in the context of the CNN architecture used. It implies that the velocity field information contained within the PLIF input is due primarily to instantaneous spatial correlations, rather than temporal dependencies. This notion requires an immense amount of further study, and will be addressed in future work.

\subsection{Decoding Unexplored Regions of the Domain}
The above analysis showcased the potency of the CNN in producing velocity fields for the same regions of the domain in which the model was trained. A natural extension of the results described above is to assess the performance of the CNN in regions of the domain that were not encountered during training. This utility of the CNN decoding could be useful in applications where one measurement is spatially restricted or blocked. 

To assess CNN performance on unexplored regions, the local CNN, denoted $\mathcal{C}_L^b$ is exclusively used (superscript $b$ indicates the box number for which the local CNN was trained). In summary, the local CNN trained on a particular box $b$ will be used to construct velocity fields in a box $q$, where $q$ is not necessarily equal to $b$. A box-dependant performance metric conditioned on $\mathcal{C}_L^b$, denoted $P_q | \mathcal{C}_L^b$, is constructed to interpret the level at which the CNN $\mathcal{C}_L^b$ is able to construct PIV fields in other boxes relative to the box it was trained on. The metric is defined as
\begin{equation}
\label{eq:metric}
     P_q | \mathcal{C}_L^b = \frac{MSE_q | \mathcal{C}_L^b - MSE_q | \mathcal{C}_L^q}{MSE_q | \mathcal{C}_L^q},
\end{equation}
where
\begin{equation}
    MSE_q | \mathcal{C}_L^b = \frac{1}{N_p}\sum_{i = 1}^{N_p} (Y_{q,i} - \mathcal{C}_{L,i}^b(X_q))^2,\quad q,b=1,...,N_b, 
\end{equation}

\begin{figure}
    \begin{center}
    \includegraphics[width=\columnwidth]{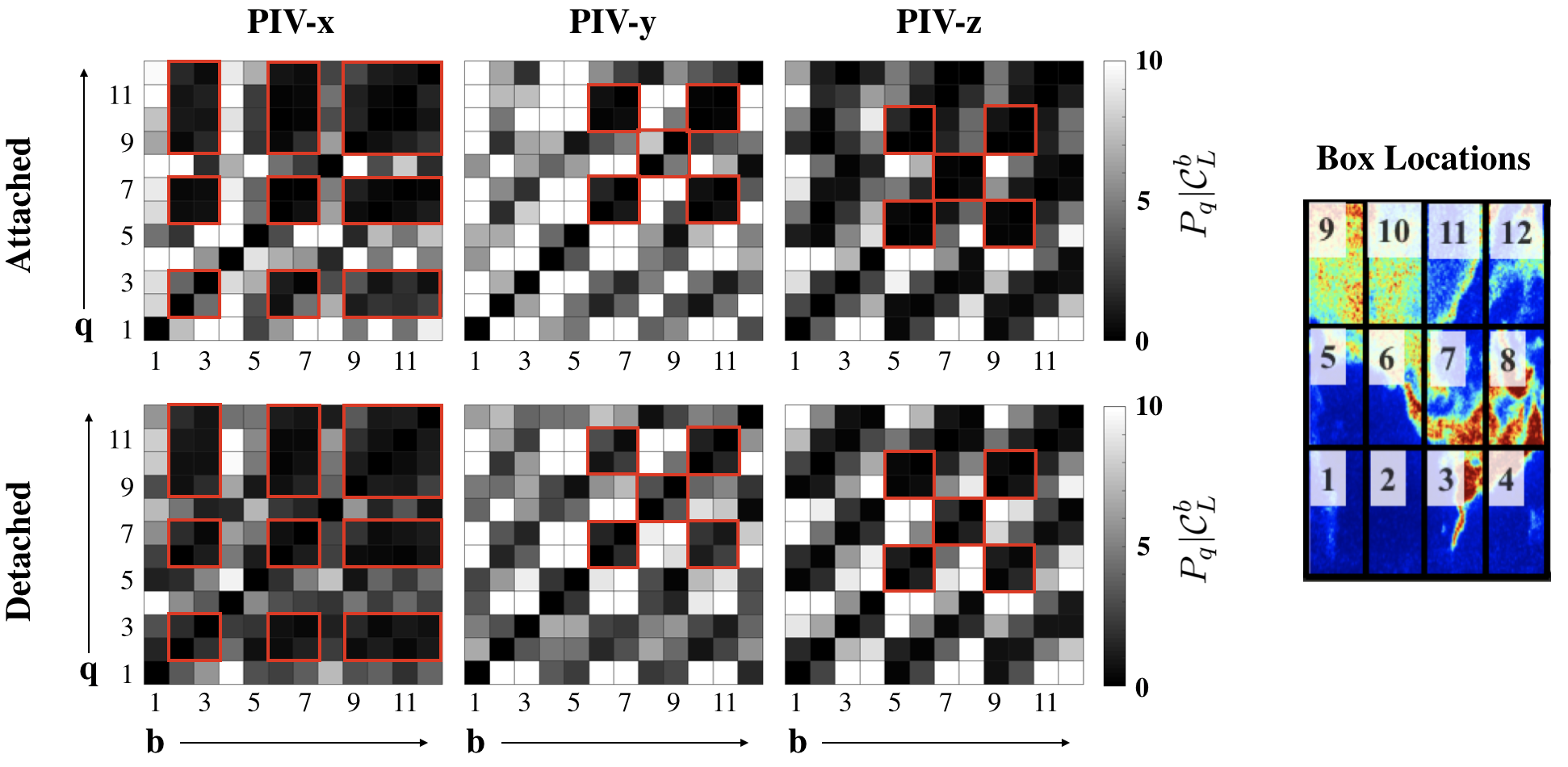}
    \caption{Values of $P_q | \mathcal{C}_L^b$ for the attached flame regime (lower is better). Columns indicate $b$ and rows indicate $q$. For example, consider b = 1 (first column) and q = 3 (third row): the corresponding location in the heatmap gives the performance measure of a local CNN trained on box 1 evaluated at box 3. On the right is an OH-PLIF image with box labels for reader convenience.}
    \label{fig:new_domain}
    \end{center}
\end{figure}

Equation~\ref{eq:metric} is a scaled MSE. The first term in the numerator is the MSE of the data in box $q$ generated from a trained model in box $b$. This quantity is scaled by the MSE of the data in box $q$ generated from a trained model \textit{in the same box $q$} (second term in numerator and denominator of Eq.~\ref{eq:metric}). This amounts to a scaling of the non-local predictions with respect to the local predictions, which gives a general measure on the extent at which domain-based extrapolation of the local CNNs can be utilized. Values of $P_q | \mathcal{C}_L^b$ organized in heatmaps are shown in Fig.~\ref{fig:new_domain} for decodings of all PIV components in the attached and detached regimes for testing set 1. Note that for cases of $q=b$ (the main diagonal of the matrices in Fig.~\ref{fig:new_domain}), the $P_q | \mathcal{C}_L^b$ values are zero. Furthermore, the matrices are not necessarily symmetric about the main diagonal. Note also that the colorbars were cut off at $P_q | \mathcal{C}_L^b = 3$ to better isolate sections of high-accuracy regions in the heat maps. 

In both attached and detached cases, the heat maps display meaningful symmetric structure (indicated by the red boxes in Fig.~\ref{fig:new_domain}. In the case of PIV-x (left-most plots), the patterns indicate that local CNNs trained on boxes in the set $b = \{2,3,6,7,9,10,11,12\}$ can be utilized to comparable accuracy on any other box in that same set. In both attached and detached cases (attached more so than detached), these specific set of boxes in the x-component of velocity is primarily dominated by velocity magnitudes close to 0 (with minor fluctuations), which explains the high amount of extrapolation accuracy. Discounting this set, the remaining boxes are $\{1,5\}$ and  $\{4,8\}$, which occupy the left and right portions of the domain, respectively. Interestingly, Fig.~\ref{fig:new_domain} implies that a model trained on box $1$ ($b=1$) is much more accurate in reporoducing PIV-x fields for boxes $5$; similarly, a model trained on box $4$ ($b=4$) is better at reproducing velocity fields in box $8$ relative to the other boxes. In contrast, a model trained on box $1$ achieves poor predictions in boxes $4$ and $8$, whereas a model trained on box $4$ achieves poor predictions in boxes $1$ and $5$. Ultimately, for boxes which capture "interesting" dynamics (largely non-zero velocity magnitudes), the trends for PIV-x in Fig.~\ref{fig:new_domain} imply low relative decoding errors for boxes on the same side of the symmetry plane as the training box and high errors for boxes on the opposite sides, indicating a capturing of antisymmetric trends characteristic of PIV-x fields. 

For both attached and detached cases, the PIV-y and z matrices exhibit pairwise checkerboard-like structures (again indicated by red boxes in Fig.~\ref{fig:new_domain}). For PIV-y, the highlighted boxes imply that models trained on boxes 6/7 can be used to good accuracy on boxes 10/11, and vice-versa. This domain extrapolation accuracy is warranted, as the set of boxes $\{6,7,10,11\}$ corresponds to the inner shear layer of largely uniform negative y-component velocity (i.e. these boxes capture similar dynamical content). Furthermore, for PIV-y, the checkerboard-like structures of dark boxes (which corresponds to a good performance metric) appears in pairs of boxes at 2/3, 6/7, and 10/11. These box pairs are symmetric about the x = 0 plane. Furthermore, a local model trained on boxes 1 and 5 can be applied to boxes 4 and 8 with reasonable relative accuracy (and vice-versa). With some exceptions, the models ultimately reflect symmetric properties of the PIV-y field about the centerline: boxes 1/4, and 5/8 enclose regions of positive y velocity, and boxes 2/3, 6/7, and 10/11 enclose regions of negative y velocity. Similar trends are present in PIV-z, except with respect to antisymmetry about the centerline (as with PIV-x) -- note that the checkerboard patterns for PIV-z are shifted one $q$ index downwards from the patterns of PIV-y, which is an object of this antisymmetry. 

Interestingly, local CNN models show the ability to 1) perform with higher relative accuracy on the same sides of the domain in the x-direction for PIV-x and PIV-z for both detached and attached flames, and 2) perform well on reflected parts of the domain in PIV-y. This potential correlation between CNN performance and flame symmetry in PIV-y, along with antisymmetry in PIV-x and z, could be used to the advantage of the user to create more accurate mappings in unexplored domain regions, and ultimately warrants a more detailed future study.

\section{Conclusions}
\label{sec:conclusion} 
A CNN was utilized to decode planar velocity fields from planar OH-PLIF images for a premixed swirl combustor in both attached and detached flame regimes. The discretization of the dataset into twelve non-overlapping boxes allowed for the comparison of global and local CNNs, both of which utilize the same underlying network architecture. Two testing datasets were utilized: one at the same operating conditions as the testing dataset, and another at entirely different operating conditions. 

For model evaluations on testing data at the same operating conditions, the velocity field decodings were much more accurate in the attached flame regime than in the detached regime in both local and global models. The higher errors in the detached regime were largely associated with inaccuracies near the combustor burner exit (boxes 2 and 3), likely caused by the  absence  of  significant  OH-PLIF  signal  in  the  presence  of  a  highly  unsteady  flame  anchoring  point driven by the precessing vortex core. Local models offered subtle improvements over the global model in terms of capturing fluctuating structures in the upper regions of the domain in the PIV-x and PIV-y components of velocity field. However, for the PIV-z field, predictions between local and global models were very similar and no noticeable improvement was provided by the local model. This implies that, since little improvement is seen in the local CNNs over the global CNN in the PIV-z decoding, the identified OH features relevant for the PIV-z mapping are much less dependent on spatial domain location (i.e. these learned features are not a function of box number for this particular component of velocity). 

Similar trends were seen for model evaluations at different operating conditions: the accuracy in the attached regime was much greater than in the detached. An encouraging (and counter-intuitive) result associated with the attached regime predictions here was a decrease in MSE values on average from the those obtained using data at the same operating conditions. In the detached regime, however, it was found that the local model generated higher errors for some portions of the domain than the global model. This implies that the detached regime at different operating conditions utilizes different physical structures in the input OH field to produce the desired velocity field outputs. On the other hand, in the attached regime, the relevant features in the OH field learned in the training phase to produce the velocity field mappings can be successfully extrapolated to different operating conditions. This means that the CNN is more ”universal” when applied to attached flames. 

In the local CNN models, the inclusion of time history in the OH field input was also analyzed. A single previous snapshot of the OH field was added to the input with the hopes of improved accuracy in the velocity field decodings. It was found that for both testing sets, the inclusion of time history led to small visible improvements in MSE in both testing sets. This has interesting implications with regards to the potentially greater physical role of instantaneous spatial correlations in the decoding over temporal dependencies. 

Lastly, the local CNN models were also used to assess the potential ability to apply a mapping learned from one section of the domain to another, previously unseen section. A scaled MSE metric was used to assess the relative local CNN performance in the various domain regions. Local CNN models showcased the ability to 1) perform well on the same sides of the domain in the x-direction for PIV-x and PIV-z, and 2) perform well on reflected parts of the domain in PIV-y; the models thus recovered properties of symmetry and anti-symmetry in the flame. 

The framework applied here brings to light avenues that can be leveraged in many experimental and modeling settings to recreate one form of data from another. Though the practical implementation indeed requires further tuning, the implications of the CNN with regards to the decoding of physically different data are very appealing. This will be studied more extensively in future work. 


\printbibliography

\section*{Appendix: Recovering High-energy Flow Directions}
The results above show model performance in decoding the velocity field in terms of direct evaluation of mean squared errors. However, other methods of interpreting model accuracy can be used based on standard time series decomposition and post-processing techniques to provide more insight to the decoding process. There are several such techniques available -- here, a proper orthogonal decomposition (POD) is conducted on the predicted velocity field time series in the attached and detached regimes to gain a different perspective on the utility of the CNN-based decoder. In short, the assessment of CNN model output based on POD allows for a direct evaluation of how well the predicted velocity field captures the distribution of variance (as well as the direction of the variance) in the true velocity field in terms of a finite set of modes, or flow directions.

First, the POD methodology is very briefly summarized. The velocity field data is recast into a set of time-averaged spatial flowfields, or modes, which form an orthonormal basis. The coordinates of the original data in this basis, called temporal coefficients, provide information on the temporal development of the flow. The velocity field data to be decomposed is denoted $\mathcal{X} \in \mathbb{R}^{N_p \times N_s}$, where $N_p$ is the number of pixels per snapshot and $N_s$ is the total number of snapshots. The orthonormal basis is contained in a matrix $U \in \mathbb{R}^{N_p \times N_m}$, where $N_m$ denotes the number of POD modes. The decomposition of each snapshot is then given by $\boldsymbol{x}_i = \sum_{j=1}^{N} a_{j,i} \boldsymbol{u}_{j}$. Here, $\boldsymbol{u}_j \in \mathbb{R}^{N_p}$ is the $j$th POD mode ($j$th column of $U$), and $a_{j,i}$ is the corresponding temporal coefficient of mode $j$ for snapshot $i$. The POD modes, which form the columns of $U$, can be found by solving the minimization problem
\begin{equation}
    \min_U ||\mathcal{X} - U U^* \mathcal{X} ||_F, \quad \text{s.t.} \quad  U^* U = I,
\end{equation}
where $*$ indicates the transpose, $F$ the Frobenius norm, and $I$ the identity matrix. This minimization problem can be solved using the singular value decomposition (SVD) of $\cal X$. To assess mode "importance", the fraction of variance captured by a projection of $\cal X$ onto mode $j$, i.e. the mode energy $\varepsilon_j$, is used. By implication of the above minimization, POD is optimal in the sense that there is no alternative set of $N_m$ modes that captures a greater amount of the variance, or energy, of $\cal X$. Mode energy is calculated directly from the singular values of $\cal X$, where 
\begin{equation}
    \label{eq:energy} 
    \varepsilon_j = \frac{\sigma_j}{\sum_k \sigma_k}.
\end{equation}
In Eq.~\ref{eq:energy}, $\sigma_j$ is the singular value associated with mode $j$ and the summation in the denominator is over all singular values of $\cal X$. 

\begin{figure}
    \centering
    \includegraphics[width =0.9\columnwidth]{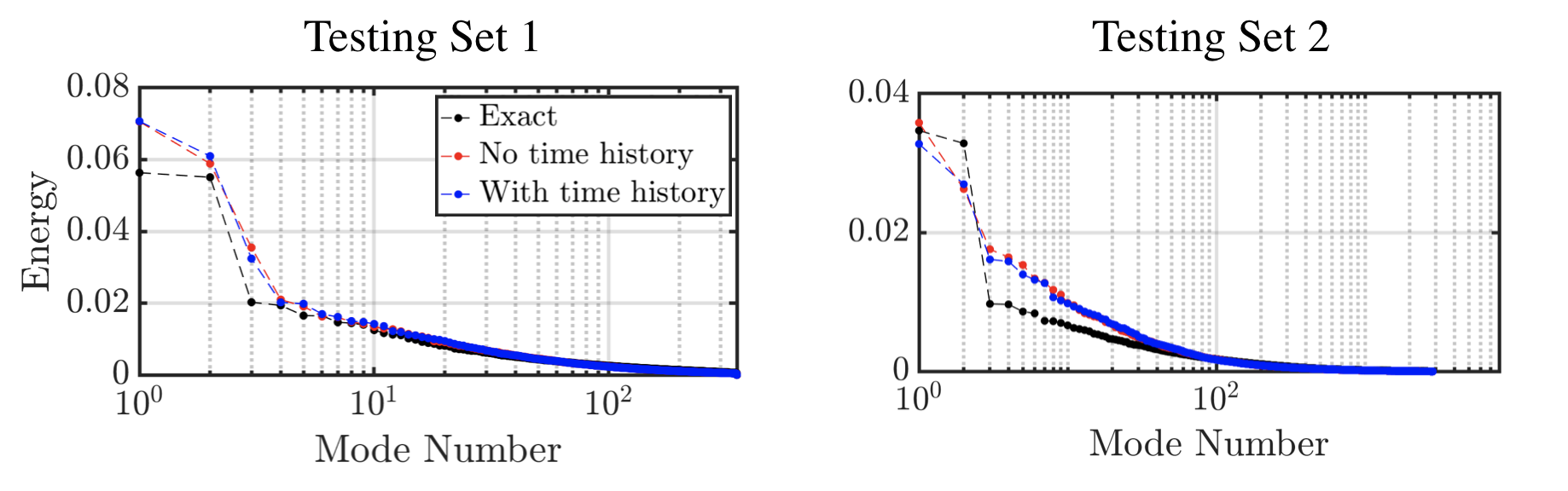}
    \caption{POD mode energies for the PIV-y component of velocity for detached flame regime shown for both testing sets. Black curve is derived from the exact, or ground-truth, time series, red curve is local CNN without time history, blue curve is with time history included.}
    \label{fig:pod_energy}
\end{figure}

For the detached flame case, POD mode energy as a function of mode number is shown for testing sets 1 and 2 in Fig.~\ref{fig:pod_energy} for the y-component of velocity field. The local CNN models with and without time history are shown (global model is omitted). In testing set 1 (same operating conditions), the predicted distribution of POD mode energy follows the exact distribution reasonably well, with slight overcompensation of mode energy into the higher modes. In testing set 2 in Fig.~\ref{fig:pod_energy}, the energy compensation of mode 1 is captured well by both models, and the predictions also generate a substantial drop-off in energy from mode 2 to mode 3. This drop-off, however, is not perfectly representative of the true curve as seen in black -- the mode energy in testing set 2 underpredicts for mode 2 and overpredicts for mode 3 onwards. Interestingly, the predicted distributions in the mode energies between the models with and without time history are very similar, despite the noticeable improvements seen with the inclusion of time history in terms of MSE seen in Sec.~\ref{sec:results_timehist}.

\begin{figure}
    \centering
    \includegraphics[width =0.6\columnwidth]{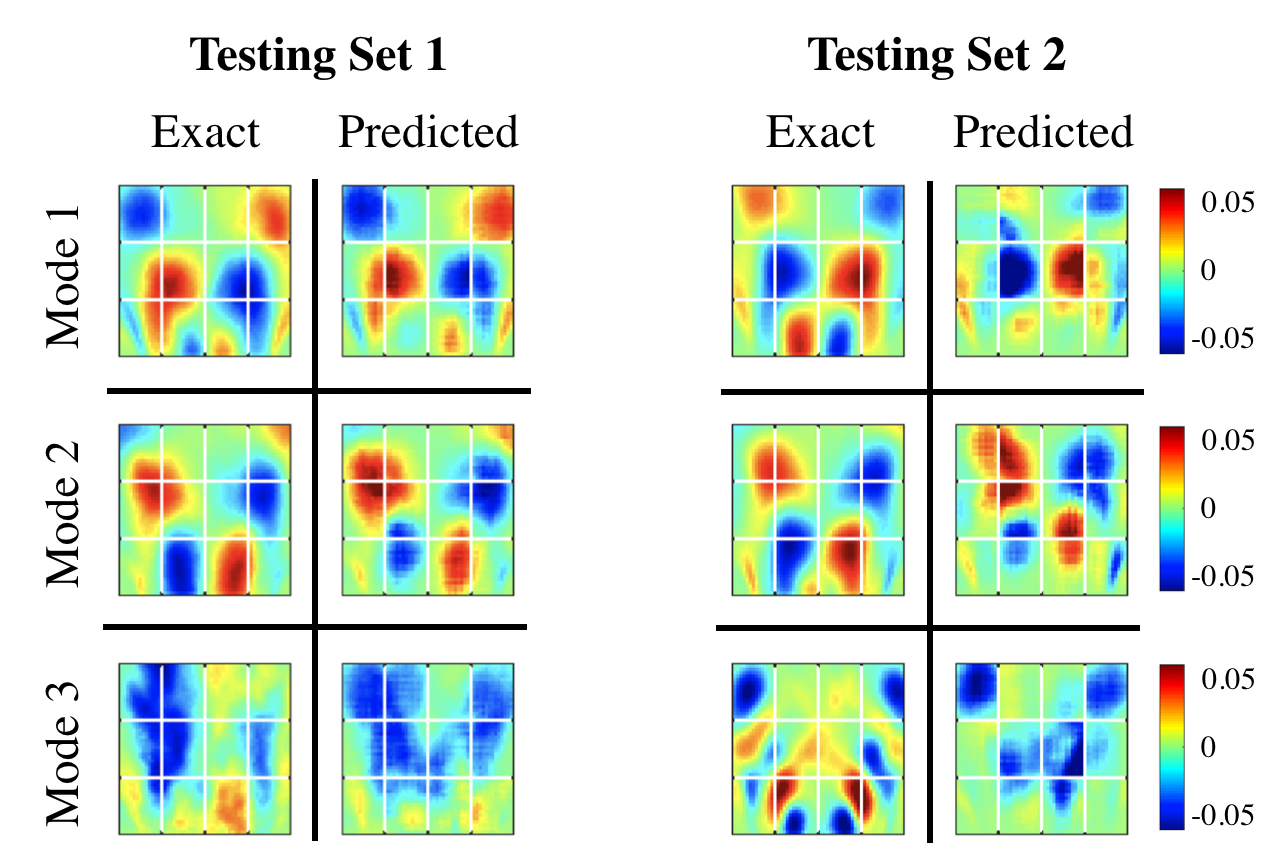}
    \caption{First three POD modes for the PIV-y component of velocity for detached flame regime shown for both testing sets. Predicted modes obtained from local CNN with time history included.}
    \label{fig:pod_modes}
\end{figure}

Alongside energies, it is important to see if the predicted mode structures are also similar in space to the exact counterparts (Fig.~\ref{fig:pod_energy} only verifies that the representation of of variance by each mode is similar, not that the modes themselves are similar). For a qualitative assessment, the first three modes from Fig.~\ref{fig:pod_energy} are shown in Fig.~\ref{fig:pod_modes} for both testing sets. Here, the predicted quantities are generated only from the local CNNs that include time history. In testing set 1, the predicted POD modes are closely capture the patterns in the exact modes with deviations occurring more in the third mode. In testing set 2, predictions are accurate for the first two POD modes, and quite inaccurate for the third mode. The model therefore does a better job in representing lower energy modes in the detached regime for unseen data at the same operating conditions, which is expected. Furthermore, in testing set 2, a noticeable discontinuity can be seen in the predicted POD mode 1 field in boxes 5 and 8 that is not present in the test set 1 counterpart. Despite this, the results shown here present the notion that the OH field contains not only information of the simultaneous velocity field, but also surprisingly accurate information regarding the high-energy flow directions in the velocity field, even for data collected from entirely different operating conditions.

\end{document}